\documentstyle[12pt,epsf]{article}
\begin{document}
\begin{center}
{\Large \bf  Diffractive hadron production and pomeron coupling
                structure

}
\bigskip
{\large S.V.~Goloskokov}
\date{}
\smallskip

{\it
Bogoliubov Laboratory of Theoretical Physics,
Joint Institute for Nuclear Research, 141980 Dubna, Russian Federation
}
\end{center}

\begin{abstract}
Large-distance effects, which lead to the spin-flip part of the
hadron -pomeron coupling in QCD -models, are discussed.
We study spin asymmetries in exclusive reactions and in
diffractive $Q \bar Q$ and vector meson production which are
sensitive to the spin-dependent part of the pomeron coupling.
\end{abstract}

Investigation of hard diffractive processes is now a problem of
topical interest. Experimental study of reactions with a
large rapidity gap \cite{h1_zeus} gives  information
on the pomeron structure. Theoretically, it is important to find
a possible way to test different model approaches which have
been proposed for the pomeron and its couplings with quarks and
hadrons. The pomeron has mainly the gluon contents and can be
represented in QCD as a two-gluon exchange \cite{low}. Thus, the
diffractive reactions may be a significant tool to study the
gluon distributions in the nucleon
at small $x$ \cite{rys}. Actually, these processes can be
expressed in terms of skewed gluon distribution in the nucleon
${\cal F}_X(X+\Delta X)$ where $X+\Delta X$ is a fraction of
the proton momentum carried by the outgoing gluon ($\Delta  X
\ll X$) and the difference between the gluon momenta
(skewedness) is equal to $X$ \cite{rad}.

The pomeron is a color singlet object which describes  high
energy reactions at fixed momentum transfer. Usually, the
pomeron exchange is written in the factorized form as a product
of the function $I\hspace{-1.6mm}P$, which absorbs $s$-
dependence of the amplitude, and the pomeron-hadron vertices
$V^{hI\hspace{-1.1mm}P}$
\begin{equation}\label{tpom} \hat T(s,t)=i I\hspace{-1.6mm}P(s,t)
V_{h_1I\hspace{-1.1mm}P} \otimes V^{h_2 I\hspace{-1.1mm}P}.
\end{equation}
The model approaches \cite{la-na, bfkl} lead to the
quark-pomeron couplings in a simple form:
\begin{equation}\label{pmu}
  V^{\mu}_{h I\hspace{-1.1mm}P} =B_{h
I\hspace{-1.1mm}P}(t)\; \gamma^{\mu},
\end{equation}
which  looks like a $C= +1$ isoscalar photon vertex
\cite{lansh-m}. In this case, the spin-flip effects are
suppressed as a power of $s$.

In  QCD-based models, which consider large-distance
contributions in hadrons, a more general form of the pomeron
coupling with the proton has been obtained. In the model
\cite{gol_mod}, which was found to be valid for momentum transfer
$|t| < \mbox{few GeV}^2$, the  $Q \bar Q$ sea effects have been
approximated by a meson cloud of the hadron. The model results in
the pomeron-hadron coupling in the form
\begin{equation}
V_{pI\hspace{-1.1mm}P}^{\mu}(p,t,x_P)
=2 p^{\mu} A(t,x_P) +\gamma^{\mu} B(t,x_P).
\label{ver}
\end{equation}
Here $x_P$ is a fraction of the initial pomeron momentum carried
by the pomeron ($x_P=0$ for elastic scattering). The
large-distance meson cloud contributions in the nucleon produce
the $A$ term in (\ref{ver}). It leads to the transverse spin
effects in the pomeron coupling which does not vanish at high
energies. This means that the pomeron might not conserve the
$s$-channel helicity. Within this model, a quantitative
description of meson-nucleon and nucleon--nucleon polarized
scattering at high energies has been obtained \cite{gol_mod}.
The model predictions for polarization at RHIC energies are shown
in Fig.1 \cite{akch}. Error bars in the Figure  indicate
expected statistical errors for the PP2PP experiment at RHIC.
The expected errors are quite small and the information about
the spin-flip part of the proton-pomeron coupling can be
obtained experimentally.

The similar structure  of the proton coupling with the
two-gluon system has been found  for moderate momentum transfer
in a QCD--based diquark model \cite{gol_kr}. Diquarks provide an
effective description of non-perturbative effects in the proton.
The spin--dependent $A$ contribution in (\ref{ver}) is
determined in the model by the effects of vector diquarks. The
predicted $A_N$ asymmetry (Fig.\ 2)  is of the same order of
magnitude as has been observed in the model \cite{gol_mod,akch}
for $|t| \sim 3 \mbox{GeV}^2$
and found in the BNL \cite{krish} and FNAL experiments
\cite{fnalp}. A similar form of the proton-pomeron coupling has
been used in \cite{schaefer}. Generally, the spin-dependent
pomeron coupling (\ref{ver}) can be obtained if one considers
together with the Dirac  the Pauli form factors \cite{nach}
in the electromagnetic nucleon current. In all the cases, the
spin-flip $A$ contribution is  determined by the
nonperturbative effects in the proton.
\\[2mm]
%%%%%%%%%%%%%%%%%%%%%%%%%%%%%%%5
\epsfxsize=14.8cm
\centerline{\epsfbox{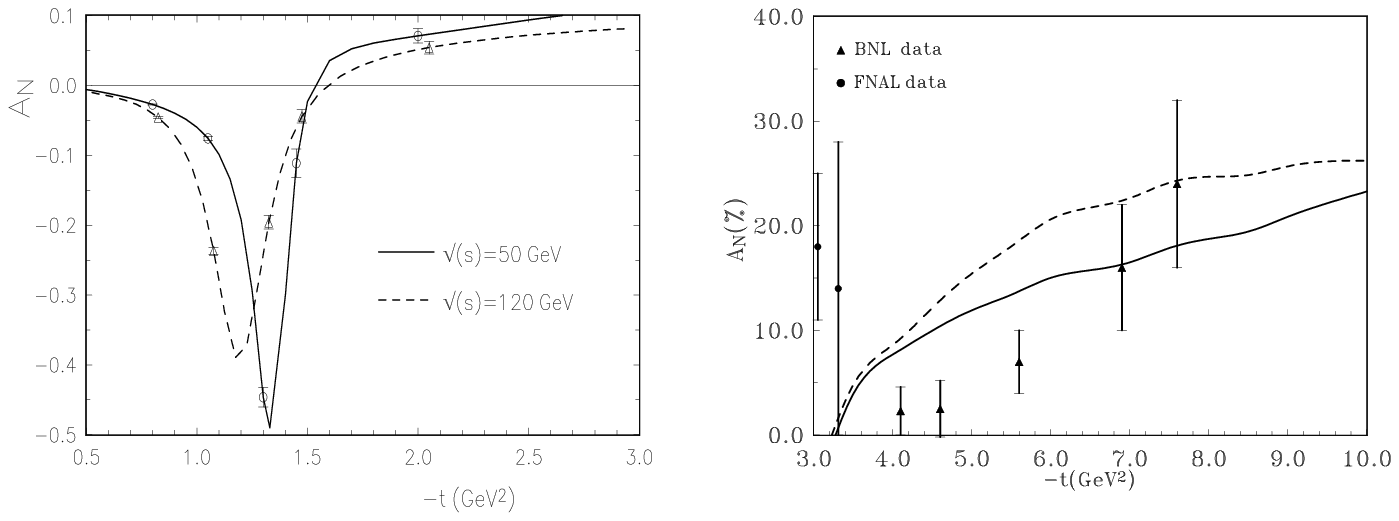}}
\\[.3cm]
%\bigskip
\noindent
\begin{minipage}{7.2cm}
Fig. 1~ { Meson cloud model predictions for single-spin
transverse asymmetry of the $pp$ scattering at RHIC energies.
}
\end{minipage}
\begin{minipage}{.15cm}
\phantom{aa}
\end{minipage}
\begin{minipage}{7.2cm}
Fig. 2~ {Diquark model predictions for single-spin asymmetry
at high energy $pp$ scattering.}
\end{minipage}
\\[2mm]
%%%%%%%%%%%%%%%%%%%%%%%%%%%%====

Let us analyze now what polarized diffractive experiments might
be sensitive to the pomeron coupling  structure (\ref{ver}). We
shall consider double spin asymmetry of the $J/\Psi$ and $Q \bar
Q$ production. The cross section of these reactions can be
decomposed into the  following important parts: leptonic and
hadronic tensors and the amplitude of the $\gamma^\star
I\hspace{-1.7mm}P \to J/\Psi (Q \bar Q)$ transition.  The hadronic tensor
for the vertex (\ref{ver}) can be written as
\begin{equation}
\label{wtenz}
W^{\mu;\nu}(s_p)= \sum_{s_{fin}} \bar u(p',s_{fin}) V_{pgg}^{\mu}(p,t,x_P)
u(p,s_p) \bar u(p,s_p)
V_{pgg}^{\star\,\nu}(p,t,x_P) u(p',s_{fin}),
\end{equation}
where $p$ and $p'$ are the initial and final proton momenta, and $s_p$
is a spin of the initial proton.

The spin-average and spin dependent cross sections with parallel
and antiparallel longitudinal polarization of a lepton and a
proton are determined by the relation
\begin{equation}
\label{spm}
\sigma(\pm) =\frac{1}{2} \left( \sigma(^{\rightarrow} _{\Leftarrow})
\pm \sigma(^{\rightarrow} _{\Rightarrow})\right).
\end{equation}
These cross sections can be expressed in terms of spin-average
and spin dependent values of the leptonic and hadronic tensors. The
structure of the leptonic tensor is well known \cite{efrem}. For the
hadronic tensor one can write
\begin{equation}
W^{\mu;\nu}(\pm)=\frac{1}{2}( W^{\mu;\nu}(+\frac{1}{2}) \pm
W^{\mu;\nu}(-\frac{1}{2})),
\end{equation}
where $W(\pm\frac{1}{2})$ are the hadronic tensors with the
helicity of the initial proton equal to $\pm 1/2$. The explicit
forms of the hadronic tensors can be found in \cite{golj}.

A simple model is considered for the amplitude of the
$\gamma^\star \to J/\Psi$ transition.  The virtual photon is
going to the $q \bar q$ state and the $q \bar q \to V$ amplitude
is described by a non-relativistic wave function
\cite{rys}. In this approximation, quarks have the same
momenta equal to half of the vector meson momentum and
$m_c=m_J/2$. The gluons from the pomeron are coupled to the
single and different quarks in the $c \bar c$ loop. This ensures
 gauge invariance of the final result.

The cross section of the $J/\Psi$ leptoproduction can be written in
the form
\begin{equation}
\label{ds}
\frac{d\sigma^{\pm}}{dQ^2 dy dt}=\frac{|T^{\pm}|^2}{32 (2\pi)^3
 Q^2 s^2 y}.
\end{equation}
For the spin-average  amplitude square we find
\begin{equation}
 |T^{+}|^2=  N ((2-2 y+y^2) m_J^2 + 2(1 -y) Q^2) s^2 [|B+2 m A|^2+|A|^2 |t|] I^2.
\label{t+}
\end{equation}
Here $N$ is a known normalization factor and $I$ is the integral
 over transverse momentum of the gluon
\begin{eqnarray}
\label{int}
I=\frac{1}{(m_J^2+Q^2+|t|)}
\int \frac{d^2l_\perp (l_\perp^2+\vec l_\perp \vec \Delta)}
{(l_\perp^2+\lambda^2)((\vec l_\perp+\vec \Delta)^2+\lambda^2)
[l_\perp^2+\vec l_\perp \vec \Delta
+(m_J^2+Q^2+|t|)/4]}.
\end{eqnarray}
The term proportional to $(2-2 y+y^2) m_J^2$ in (\ref{t+})
represents the contribution of the virtual photon  with
transverse polarization. The $2(1 -y) Q^2$ term describes the
effect of longitudinal photons.

The spin-dependent amplitude square looks like
\begin{equation}
|T^{-}|^2= N (2- y)  s |t|  [|B|^2+ m (A^\star B +A B^\star)] m_J^2 I^2.
\label{t-}
\end{equation}
As a result, we find the following form of asymmetry \cite{golj}:
\begin{equation}
\label{asy}
A_{ll} =\sigma(-)/\sigma(+) \sim \frac{|t|}{s}
\frac{(2-y) [|B|^2+ m (A^\star B+ A B^\star)]}
{(2-2 y+ y^2)[|B+2 m A|^2+ |t| |A|^2]}.
\end{equation}

The $A_{ll}$ asymmetry of vector meson production is equal to
zero for the forward direction ($t=0$). It depends on the ratio
of the spin-flip to the non-flip  parts of the pomeron coupling
$\alpha=A/B$. The absolute value of $\alpha$ is proportional
to the ratio of helicity-flip and non-flip amplitudes which have
been found in \cite{gol_mod,gol_kr} to be of about 0.1 and
 weakly dependent on energy. The predicted asymmetry at HERMES
energies is shown in Fig.\ 3. At HERA energies, the asymmetry will
be negligible. The value of asymmetry for $\alpha=0$ is not
equal to zero. This term of the asymmetry is determined by the
$\gamma_\mu$ part of the pomeron coupling (\ref{ver}). It gives
the predominated contribution to the asymmetry of vector meson
production in our model.
\\[1mm]
%%%%%%%%%%%%%%%%%%%%%%%%%%%%%%%5
\begin{minipage}{7.1cm}
\phantom{.}
\vspace{1.5mm}
\epsfxsize=6.8cm
\centerline{\epsfbox{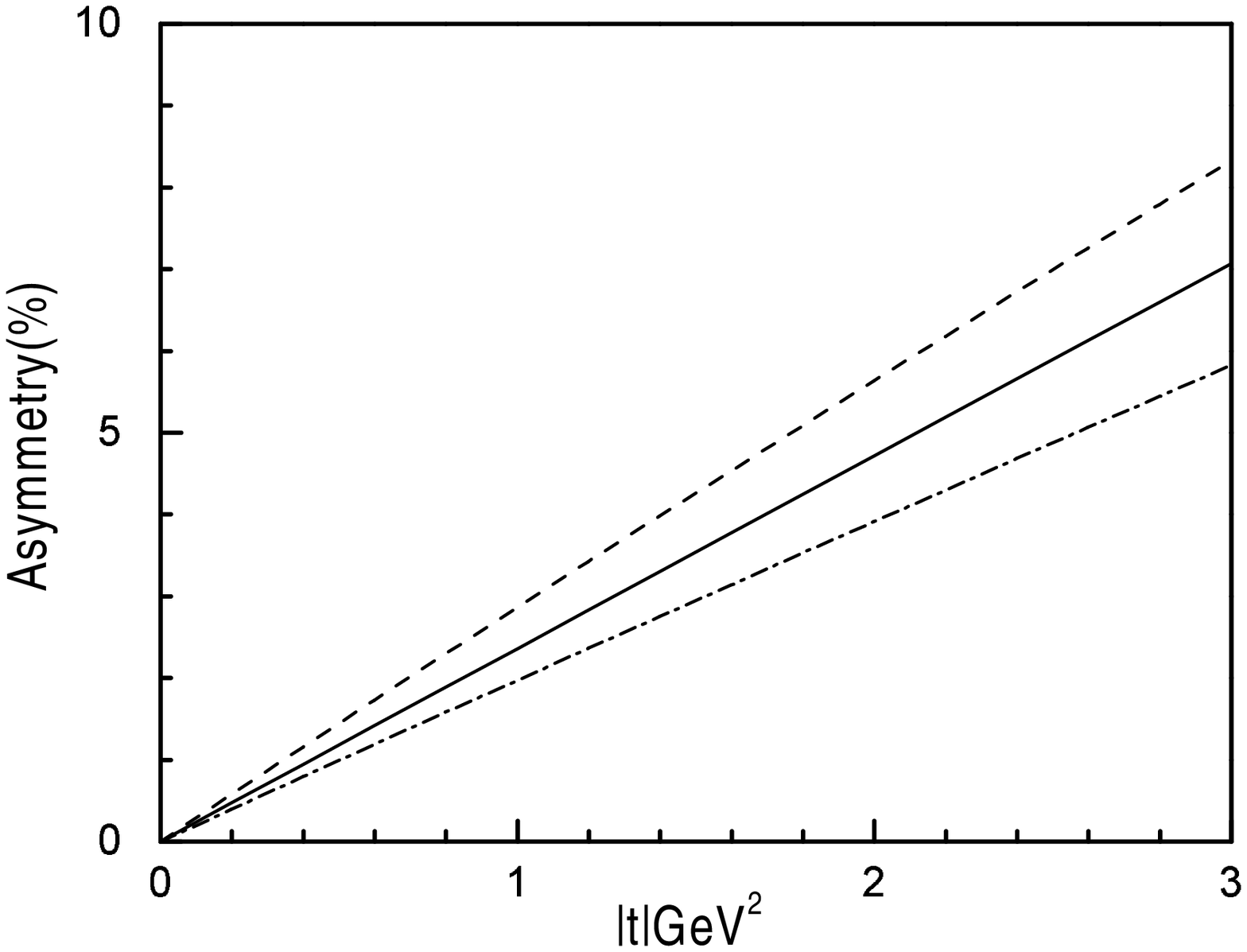}}
\end{minipage}
\begin{minipage}{0.12cm}
\phantom{aa}
\end{minipage}
\begin{minipage}{7.1cm}
\epsfxsize=8.cm
\centerline{\epsfbox{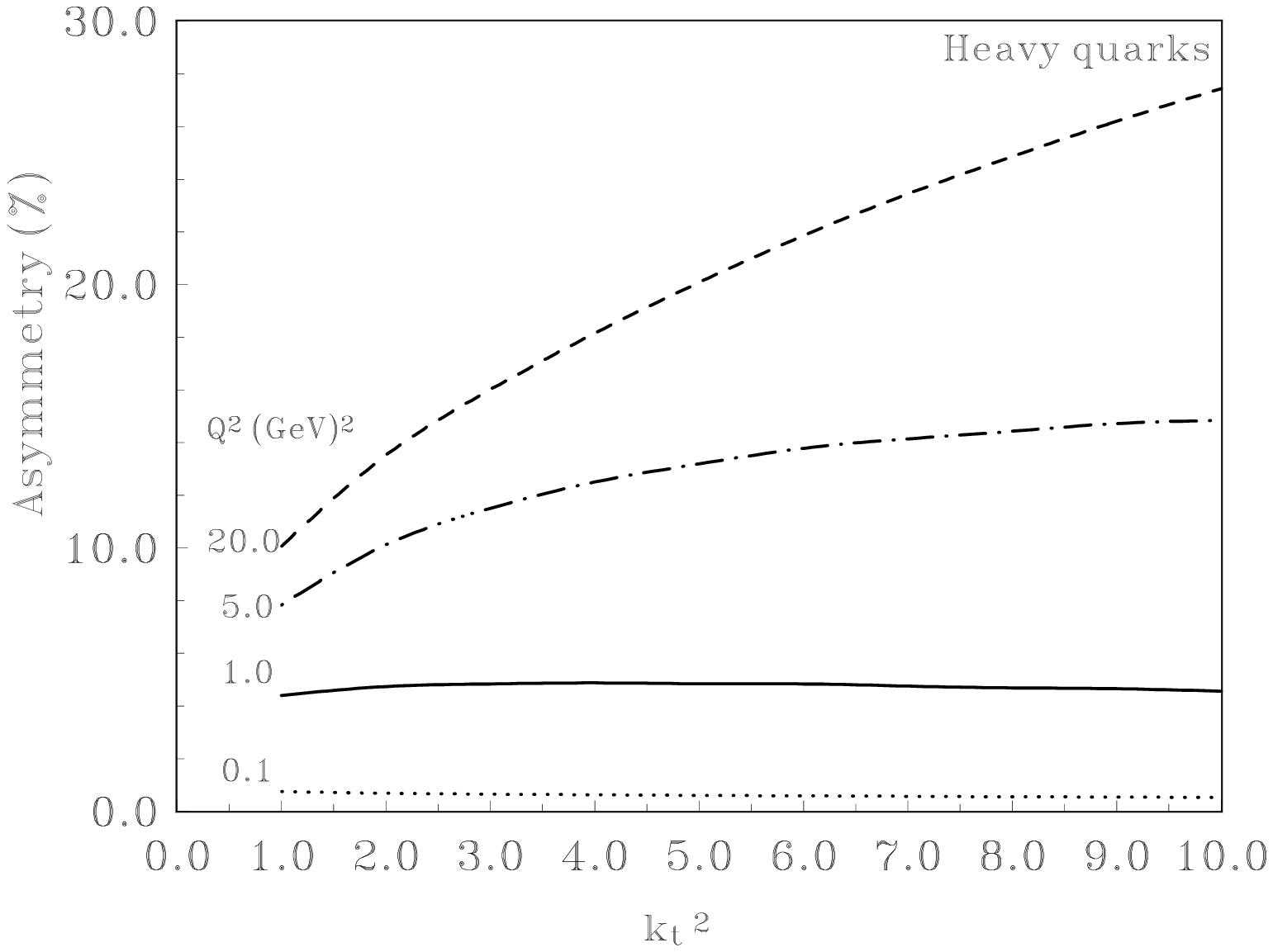}}
\end{minipage}
\medskip
\begin{minipage}{7.2cm}
Fig.3~ { The predicted $A_{ll}$ asymmetry of the $J/\Psi$ production
at HERMES: solid line -for $\alpha=0$; dot-dashed and dashed lines -for
$\alpha=\pm 0.1\mbox{GeV}^{-1}$.}
\end{minipage}
\begin{minipage}{.15cm}
\phantom{aa}
\end{minipage}
\begin{minipage}{7.2cm}
Fig.4~ { The predicted $Q^2$ dependence of $A_{lT}$ asymmetry
for the $c \bar c$ production at HERA  for
$\alpha=0.1\mbox{GeV}^{-1}$, $x_p$=0.1, $y$=0.5}.
\end{minipage}
%%%%%%%%%%%%%%%%%%%%%%%%%%%%====
\\[2mm]

Let us pass now to spin effects in $Q \bar Q$ leptoproduction.
In the two-gluon picture of the pomeron, we consider all the
graphs where the gluons from the pomeron couple to a different
quark as well to the single one. The spin-average
and spin-dependent cross section can be written in the form
\begin{equation}
\label{sigma}
\frac{d^5 \sigma(\pm)}{dQ^2 dy dx_p dt
dk_\perp^2}=
\left(^{(2-2 y+y^2)} _{\hspace{3mm}(2-y)}\right)
 \frac{N(x_p,Q^2) \; C(\pm)}
{\sqrt{1-4k_\perp^2\beta/Q^2}}.
\end{equation}
Here, $N(x_p,Q^2)$ is a normalization function which is common
for spin average and spin dependent cross section and
\begin{eqnarray}
\label{tpm}
C(\pm) = \int \frac{ d^2 l_{\bot} d^2 l_{\bot}'
D^{\pm}(t,Q^2,l_{\bot},l'_{\bot},\cdots)}
{(l_\perp^2+\lambda^2)((\vec l_\perp+\vec
r_{\perp})^2+\lambda^2) (l_\perp^{'2}+\lambda^2)((\vec l'_\perp+\vec
r_{\perp})^2+\lambda^2)},
\end{eqnarray}
where $D^{\pm}$ function comprises a sum  of  the $\gamma P \to
Q \bar Q$ production diagrams and the corresponding crossed
contributions convoluted with the spin average and
spin-dependent tensors.
The obtained diffractive $A_{ll}$ asymmetry has weak energy
dependence and is proportional to $x_p$ which is typically of
about $.05-.1$. The predicted asymmetry is quite small and does
not exceed 1-1.5\% \cite{goldesy}. We find that the asymmetry
is not equal to zero for $\alpha=0$. The value of the asymmetry for
nonzero $\alpha$ is determined by the spin--dependent part of
the pomeron coupling. However, as in the case of $J/\Psi$
production, sensitivity of the asymmetry to $\alpha$ is not
very strong.

Another object, which can be studied at polarized $Q \bar Q$
production, is the $A_{lT}$ asymmetry with longitudinal lepton
and transverse proton polarization. It has been found that the
$A_{lT}$ asymmetry is proportional to the scalar production of
the proton spin vector and the transverse jet momentum. Thus,
the asymmetry integrated over the azimuthal jet angle is zero.
We have calculated the $A_{lT}$ asymmetry for the case when the
proton spin vector is perpendicular to the lepton scattering
plane and the jet momentum is parallel to this spin vector.  The
estimated $Q^2$ dependence of the $A_{lT}$ asymmetry integrated
over $t$ for $\alpha=0.1\mbox{GeV}^{-1}$ is shown in Fig. 4. The
predicted asymmetry is huge and has a strong $k^2_{\perp}$
dependence. The large value of $A_{lT}$ asymmetry is caused by
the fact that it does not have a small factor $x_p$ as a
coefficient.

In the present report, the polarized cross section of the
diffractive hadron leptoproduction at high energies has been
studied. The spin asymmetries are expressed in terms of the $A$ and
$B$ structures of the pomeron coupling (\ref{ver}). Generally,
the function $B$ should be determined by the spin--average and
the function $A$ - by the polarized skewed gluon distribution in
the proton. The $B \gamma^{\mu}$ term of the pomeron coupling
(\ref{ver}) contributes to both $\sigma(+)$ and $\sigma(-)$ cross
sections which for $\alpha=0$ are proportional to $B^2$. This
gives the nonzero $A_{ll}(\alpha=0)$ asymmetry which is
independent of the gluon density. We predict not small value of
the $A_{ll}$ asymmetry of the diffractive vector meson
production at the HERMES energy.  The obtained asymmetry is
independent of the mass of a produced meson. So, we can expect a
similar value of the asymmetry in the polarized diffractive
$\phi$ --meson leptoproduction. The predicted $A_{ll}$ asymmetry
in the $Q \bar Q$ leptoproduction is smaller than 1.5\%. The
$A_{ll}(\alpha=0)$ contribution is predominated in asymmetry, and
sensitivity of the asymmetry on $\alpha$ for $\alpha \neq 0$  is
rather weak. Thus, the $A_{ll}$ asymmetry in diffractive
reactions is not a good tool to study polarized gluon
distributions of the proton and the spin structure of the pomeron.
Otherwise, it has been found not small $A_{lT}$ asymmetry in
diffractive $Q \bar Q$ production. This asymmetry is
proportional to $\alpha$ and can be used to obtain direct
information about the spin--dependent part $A$ of the pomeron
coupling. Experimental analyses of energy dependence of the $A_{lT}$
asymmetry as well as of the $A_{N}$ asymmetry in elastic $pp$
scattering, which have a weak energy dependence in the model, can
throw light on the spin structure of the pomeron coupling. They are
appropriate objects to study the polarized gluon structure of
the proton too. Thus, the pomeron coupling structure can be
investigated in diffractive processes. This gives important
information on the spin structure of QCD at large distances.


\begin{thebibliography}{99}
\bibitem{h1_zeus} ZEUS Collaboration,  M.\ Derrick et al.,  Z.\ Phys.
{\bf C68}, 569 (1995);\\
H1 Collaboration,  T.\ Ahmed et al., Phys.\ Lett. {\bf B348},
 681 (1995).
\bibitem{low} F.E.\ Low, Phys.\ Rev.\ {\bf D12}, 163 (1975); \\
              S.\ Nussinov, Phys.\ Rev.\ Lett.\ {\bf 34}, 1286 (1975).
\bibitem{rys} M.G. Ryskin, Z.Phys. {\bf C57}, 89 (1993);\\
S.J. Brodsky at al., Phys.Rev. {\bf D50}, 3134 (1994).
\bibitem{rad} A.V.\ Radyushkin, Phys.\ Rev.\ {\bf D56}, 5524 (1997);\\
  X.\ Ji,  Phys.\ Rev.\ {\bf D55}, 7114 (1997).
\bibitem{la-na} P.V.\ Landshoff, O.\ Nachtmann,  Z.\ Phys. {\bf C35}, 405 (1987).
\bibitem{bfkl} E.A.\ Kuraev, L.N.\ Lipatov, V.S.\ Fadin, Sov.\ Phys.\ JETP
{\bf 44}, 443 (1976);
\bibitem{lansh-m} A.\ Donnachie, P.V.\ Landshoff,
                  Nucl.\ Phys.\ {\bf B244}, 322 (1984).
\bibitem{gol_mod}  S.V.\ Goloskokov, S.P.\ Kuleshov, O.V.\ Selyugin,
           Z.\ Phys. {\bf C50},  455 (1991).
\bibitem{akch}  N. Akchurin, S.V. Goloskokov, O.V. Selyugin,
Int. J. Mod. Phys. {\bf A14}, 253 (1999).
\bibitem{gol_kr} S.V.\ Goloskokov, P.\ Kroll, Phys.\ Rev. {\bf D60}, 014019 (1999).
\bibitem{krish} D.C.\ Peaslee et al.  Phys.\ Rev.\ Lett.\ {\bf 51}, 2359
(1983).
 \bibitem{fnalp} G.\ Fidecaro at al., Phys.\  Lett.\ {\bf B105}, 309
(1981).
\bibitem{schaefer} J.\ Klenner, A.\ Sch\"afer, W.\ Greiner, Z.\ Phys. {\bf A352}, 203
(1995).
\bibitem{nach} T.\ Arens, M.\ Diehl, O.\ Nachtmann, P.V.\ Landshoff,
     Z.\ Phys. {\bf C74}, 651 (1997).
\bibitem{efrem}  M.\ Anselmino, A.\ Efremov, E.\ Leader, Phys.\
Rep. {\bf C261}, 1 (1995).
\bibitem{golj} S.V.\ Goloskokov, Eur. Phys. J. {\bf C11}, 309 (1999).
\bibitem{goldesy} S.V.\ Goloskokov, to appear in Proc. of the Workshop  "Polarized Protons
at High Energies - Accelerator Challenges and Physics Opportunities" (Hamburg, Germany, 17-20 May
1999), e-Print: hep-ph/9907429.
\end{thebibliography}
\end{document}